\documentclass[reprint,aps,pra,superscriptaddress,showpacs,
amsmath,amssymb]{revtex4-1}
\usepackage{mathtools}
\usepackage{graphicx}
\usepackage{booktabs,multirow}

\begin{document}
\newcommand{\pd}[2]{\frac{\partial #1}{\partial #2}} 
\newcommand{\ket}[1]{\left| #1 \right>} 
\newcommand{\bra}[1]{\left< #1 \right|} 
\renewcommand{\b}{\mathbf}

\title{Manipulation of optical-pulse-imprinted memory in a $\Lambda$ system}

\author{Rodrigo Guti\'{e}rrez-Cuevas}
\email{rgutier2@ur.rochester.edu}
\affiliation{Center for Coherence and Quantum Optics, University of 
Rochester, Rochester, New York 14627, USA}
\affiliation{Institute of Optics, University of Rochester, Rochester, New 
York 14627, USA}
\author{Joseph H. Eberly}
\affiliation{Center for Coherence and Quantum Optics, University of 
Rochester, Rochester, New York 14627, USA}
\affiliation{Department of Physics and Astronomy, University of Rochester, 
Rochester, New York 14627, USA}

\date{\today}

\begin{abstract}
We examine coherent memory manipulation in a $\Lambda$-type medium, using the 
 second order solution presented by Groves, Clader and Eberly [J. Phys. 
B: At. Mol. Opt. Phys. \textbf{46}, 224005 (2013)] as a guide. The analytical 
solution obtained using the Darboux transformation and a nonlinear 
superposition principle describes complicated soliton-pulse dynamics which, 
by an appropriate choice of parameters, can be simplified to a well-defined 
sequence of pulses interacting with the medium. In this report, this solution 
is reviewed and put to test by means of a series of numerical simulations, 
encompassing all the parameter space and adding the effects of homogeneous 
broadening due to spontaneous emission. We find that even though the 
decohered results deviate from the analytical prediction they do follow a 
similar trend that could be used as a guide for future experiments. 
\end{abstract}


\maketitle

\section{Introduction}

The seminal work by McCall and Hahn \cite{mccall1967self,mccall1969self} 
showed the relevance of a semiclassical treatment of light-matter 
interactions for strong fields with intensities far above the one-photon 
limit. In this regime, disagreements with quantum electrodynamics are not 
noticeable. Their discovery of self-induced transparency (SIT) showed that 
new kinds of interactions beyond the well-known Beer's law were possible. 
This paved the way for a number of interesting phenomena such as coherent 
population trapping \cite{gray1978coherent}, electromagnetically induced 
transparency (EIT) 
\cite{kasapi1995electromagnetically,harris2008electromagnetically}, and slow 
and fast light \cite{garrett1970propagation,boyd2002slow,milonni2004fast}, to 
mention a few. There have been a number of discussions dealing with the 
validity of the semiclassical theory against a full QED treatment (see for 
example \cite{milonni1976semiclassical}), but no one argues about its 
utility. Even today we continue to reap the benefits from this ``incomplete'' 
theory.

Some of these phenomena have been used to achieve light storage and 
manipulation \cite{grobe1994formation}. Light can be slowed up to the point 
where it stops and is stored in the medium \cite{fleischhauer2000dark}. 
Then it can be regenerated as was observed in \cite{liu2001observation}. 
Some other schemes have been employed such as a 
combination of EIT and four-wave mixing in hot atomic vapor 
\cite{camacho2009four}. The main potential application of the storage and 
retrieval of light is towards quantum memories. Quantum optical systems are 
desirable for this purpose as they have small decoherence and short 
interaction times \cite{milonni2004fast}. Here we test the fidelity of the 
complicated atom-pulse dynamics given by the second order solution 
derived in \cite{groves2013jaynes} to one of the major sources of 
decoherence: spontaneous emission.

\begin{figure}
\includegraphics[scale=1]{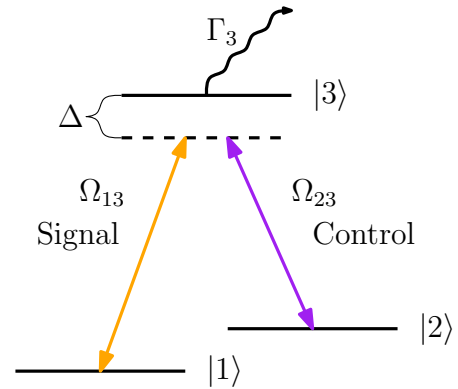}
\caption{\label{lambda} (Color online) Three-level
 atom in a $\Lambda$ configuration 
interacting with two fields in two-photon resonance via the common detuning 
$\Delta$, with spontaneous emission $\Gamma_3$ from the excited state.}
\end{figure}

The interaction of strong electromagnetic fields with atomic systems leads to 
nonlinear dynamics, which makes it difficult to solve analytically, but it is 
worth the effort. The SIT solution to the Maxwell-Bloch equation for a 
two-level atom made clear the importance of the pulse area for the 
interaction. This is defined as
\begin{equation}
\theta(x,t)=\int^t_{-\infty}\Omega(x,t)dt.
\label{area}
\end{equation}
When one takes the limit of infinite time we get the entire area of the given 
pulse which follows the predictions of the area theorem, namely, the pulse 
area tends to the closest even multiple of $\pi$. This results from the 
smoothing properties of Doppler broadening \cite{allen2012optical} by taking 
the average over the corresponding inhomogeneous distribution function of the 
atomic part in the evolution equation for the field. Here we further explore 
the usage of nonlinear optical interaction for light storage and memory 
manipulation in a $\Lambda$-type system (see Fig.~\ref{lambda}) of ultra cold 
atoms, where it is appropriate to neglect the effects of collisional and 
Doppler broadening.   

\section{Mathematical Model}

We consider the interaction of two fields with a $\Lambda$-system in 
two-photon resonance with each field addressing a different atomic transition 
as shown in Fig.~\ref{lambda}.
Each field interacts with the atomic system via the dipole moment operator 
which only links levels 1 to 3 and 2 to 3; 
$\hat{\vec{d}}=\vec{d}_{13}\ket{1}\bra{3}+\vec{d}_{23}\ket{2}\bra{3}
+\vec{d}_{31}\ket{3}\bra{1}+\vec{d}_{32}\ket{3}\bra{2}$.
We write the fields in carrier-envelope form:
\begin{align}
\vec{E}(x,t)=&\vec{\mathcal E}_{13}(x,t)e^{i(k_{13}x
-\omega_{13}t)}\nonumber\\
&+\vec{\mathcal E}_{23}(x,t)e^{i(k_{23}x-\omega_{23}t)}+c.c
\label{carrier}
\end{align}
where $\omega_{13}$ and $\omega_{23}$ are the field frequencies, $k_{13}$ and 
$k_{23}$ the vacuum wave numbers and $\vec{\mathcal E}_{13}(x, t )$ and 
$\vec{\mathcal E}_{23}(x,t)$ the slowly-varying field envelopes. We 
assume that the envelopes change slowly over many cycles of the optical 
frequency, thus justifying the slow-varying envelope approximation (SVEA). 
Following \cite{groves2013jaynes} we refer to the 1-3 field as the signal 
pulse and the 2-3 field as the control pulse. In the rotating-wave 
approximation (RWA) the bare frequencies $\omega_{13}$ and $\omega_{23}$ are 
eliminated in favor of $\Delta$, their common detuning, and the total 
Hamiltonian takes the form:
\begin{equation}
\hat H=-\frac{\hbar}{2} \left(
\begin{array}{ccc} 
0&0&\Omega_{13}^*\\
0&0&\Omega_{23}^*\\
\Omega_{13}&\Omega_{23}&-2\Delta
\end{array}
\right),
\label{hrwa}
\end{equation}
where we defined the Rabi frequencies, $\Omega_{13}(x,t)=2\vec{d}_{31}\cdot 
\vec{\mathcal E}_{13}(x, t )/\hbar$ and $\Omega_{23}(x,t)=2\vec{d}_{32}\cdot 
\vec{\mathcal E}_{23}(x, t )/\hbar$, and the detuning $\Delta=(E_3-
E_1)/\hbar-\omega_{13}=(E_3-E_2)/\hbar-\omega_{23}$ ($E_i$ corresponds to the 
energy of level $\ket i$). The dynamics of the system are dictated by the von 
Neumann equation for the density matrix of the atomic sample:
\begin{equation}
i\hbar \pd{\hat \rho}{t}=[\hat H,\hat \rho],
\label{neumann}
\end{equation}
and by Maxwell's wave equation in the SVEA for the field evolution:
\begin{subequations}
\label{meqs}
\begin{align}
\left(\pd{ }{x}+\frac{1}{c}\pd{}{t}\right)\Omega_{13}&=i\mu_{13} \rho_{31}\\
\shortintertext{and}
\left(\pd{ }{x}+\frac{1}{c}\pd{}{t}\right)\Omega_{23}&=i\mu_{23} \rho_{32}.
\end{align}
\end{subequations}
Here, we defined the atom-field coupling parameters $\mu_{j3}=N\omega_{j3}|
d_{j3}|^2/\hbar \epsilon_0c$ with $j=1,\,2$. 

We consider the case of coherent short pulses for which it is justified to 
neglect homogeneous relaxation processes due to the fast interaction with the 
medium. This gives us a set of eight nonlinear partial differential equations 
that need to be solved simultaneously. As has been shown by Park and Shin 
\cite{park1998matched} and Clader and Eberly \cite{clader2007two}, for the 
case of two-photon resonance and equal atom-field coupling parameters, 
$\mu_{13}=\mu_{23}=\mu$, the system of equations given by Eqs.~\eqref{neumann} 
and  
\eqref{meqs} become integrable and thus can be solved by methods such as 
inverse scattering \cite{lamb1980elements}, the B\"acklund transformation 
\cite{miura1976backlund}, and the Darboux transformation 
\cite{gu2006darboux,cieslinski2009algebraic}. This can be easily shown by the 
introduction of the constant matrix
\begin{equation}
\hat W=i\ket{3}\bra 3 = \left(
\begin{array}{ccc}
0&0&0\\
0&0&0\\
0&0&i
\end{array}
\right),
\end{equation}
so that Eqs.~\eqref{neumann} and \eqref{meqs} in the traveling-wave 
coordinates $T=t-x/c$ and $Z=x$ can be expressed as:
\begin{subequations}
\label{mb}
\begin{equation}
\label{neumann2}
i\hbar \pd{\hat \rho}{T}=[\hat H,\hat \rho] 
\end{equation}
and
\begin{equation}
\pd{\hat H}{Z}=-\frac{\hbar \mu}{2}[\hat W,\hat \rho].
\end{equation}
\end{subequations}

By combining these two equations it is clear that the Lax equation, 
\begin{equation}
\partial_Z \hat U-\partial_T \hat V +[\hat U,\hat V]=0,
\end{equation}
is satisfied where the Lax operators are defined as $\hat U=-(i/\hbar)\hat H 
- \lambda \hat W$ and $\hat V=(i\mu/2\lambda)\hat \rho$, and $\lambda$ is a 
constant known as the spectral parameter. This effectively shows that the 
Maxwell-Bloch equations \eqref{mb} are integrable.

\begin{figure}
\includegraphics[scale=0.9]{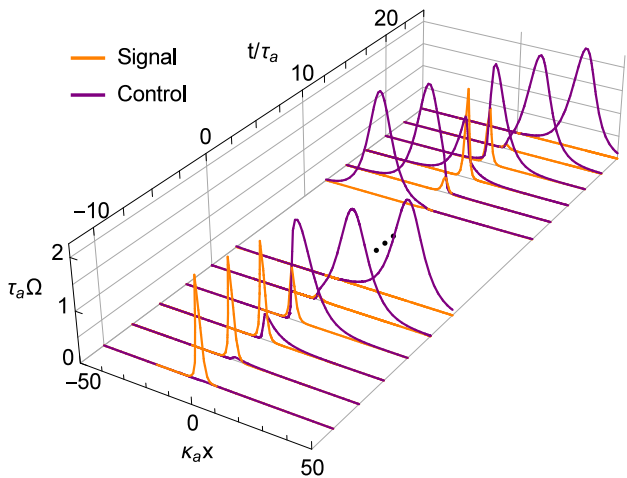}
\caption{\label{pulses} (Color online)Pulse evolution dictated by the 
second order analytical solution obtained by Groves \emph{et al.} 
\cite{groves2013jaynes}. Appropriate parameters were chosen so that the 
intricate analytical solution could be simplified into a well-defined 
sequence of pulses, composed of two steps (separated by the ellipsis). In the 
first step, at $t/\tau_a=-10$, a $2\pi$-signal pulse comes in; and as it 
interacts with the medium, it gives way to a control pulse while storing its 
information at $\kappa_a x_1=0$. During the storage process, which takes 
place between $t/\tau_a=-7.5$ and $2.5$, the areas of the individual pulses 
are no longer equal to $2\pi$ but the total pulse area [see Eq.~\eqref{atot}] 
remains constant and equal to $2\pi$. At $t/\tau_a=2.5$ we can see that the 
initial $2\pi$ signal pulse is gone and has been taken over by a $2\pi$
control pulse propagating away at the speed of light. The second step starts 
at $t/\tau_a=12.5$ as a $2\pi$ control pulse comes in and collides with the 
information imprint left by the signal pulse. During the collision, which 
takes place between times $t/\tau_a=15$ and $22.5$, the initial signal pulse 
is retrieved and redeposited at $\kappa_a x_2=3$, effectively pushing the 
imprint farther into the medium. Here again, during the interaction the total 
pulse area is conserved and equal to $2\pi$. When the re-encoding has taken 
place, the control pulse recovers its original pulse area of $2\pi$ and 
propagates away at the speed of light. The corresponding imprint is depicted 
in Fig.~\ref{density}.}
\end{figure}

The solution obtained in \cite{groves2013jaynes} is a second-order solution 
obtained from the nonlinear superposition of two first-order solutions, which 
in turn were obtained by means of the Darboux transformation from the trivial 
solution of a quiescent medium, $\hat \rho=\ket 1 \bra 1 $, and no fields, 
$\Omega_{13}=\Omega_{23}=0$. With an appropriate choice of parameters, this 
complicated solution can be reduced to a well-defined sequence of pulses 
interacting with the medium (see Fig.~\ref{pulses}), transferring information 
back and forth. 

As was done in \cite{clader2007two} we will define the total pulse area as 
\begin{equation}
\theta_\text{tot}=\sqrt{|\theta_{13}|^2+|\theta_{23}|^2},
\label{atot}
\end{equation}
and we have that $\theta_\text{tot}=2\pi$ for the two first-order solutions. 
This concept can be extended to the second-order solution by applying it to 
the pulses in each step of the sequence, as long as they are sufficiently 
separated, as in Fig.~\ref{pulses} where the two steps are separated by an 
ellipsis. 
We will label the parameters pertaining to different steps of the sequence of 
pulses by the letters $a$ and $b$.
For the first step, in the limit $t/\tau_a\ll-1$ we have a SIT-like signal 
pulse propagating, driving population from the ground state $\ket 1$ into the 
excited state $\ket 3$ and coherently driving it back, thus obtaining the 
characteristic $2\pi$ pulse shaped as an hyperbolic secant. As the control 
pulse is only zero in the limit of infinite negative time, some of the 
excited population is coherently driven into the ground state $\ket 2$, thus 
amplifying the seed of the control pulse and depleting the signal pulse. 
During this transfer the signal pulse encodes its information into the 
ground-state elements $\rho_{11}$, $\rho_{22}$, and $\rho_{12}$ of the density 
matrix. This encoding we refer to as an imprint. After the storage process is 
over we get a $2\pi$ control pulse propagating away at the speed of light as 
it is decoupled from the medium. Both signal and control pulses have a 
duration of $\tau_a$, and they are time-matched. Therefore the ratio between 
their Rabi frequencies is independent of time and given by:
\begin{equation}
\frac{\Omega_{13}^a(x,t)}{\Omega_{23}^a(x,t)}=e^{-\kappa_a( x- x_1)},
\end{equation}
where the absorption coefficient is given by $\kappa_a=\mu\tau_a/2$, since we 
are neglecting the effects of Doppler broadening, and $x_1$ is the location 
of the imprint. This relation shows us how we should map the analytical 
solution to appropriate initial conditions for the numerical computation. It 
is easy to see that, by integrating the previous equation with respect to $T$ 
and by considering that our medium starts at $x=0$, we can get $x_1$ in terms 
of the pulse areas:
\begin{equation}
\kappa_a x_1=\ln \left(\frac{\theta_{13}^a(x=0)}{\theta_{23}^a(x=0)}\right).
\label{ximp}
\end{equation}

\begin{figure}[b]
\includegraphics{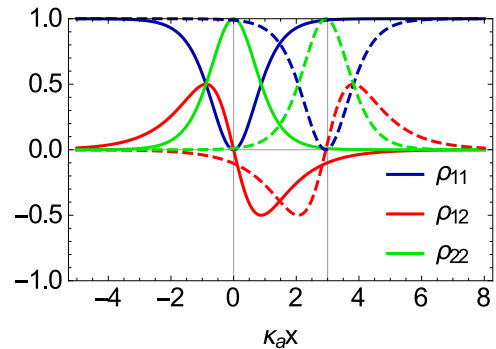}
\caption{\label{density} (Color online) Information 
encoding into a $\Lambda$-system: 
Imprint as it has been encoded in the ground state density matrix elements 
before (continuous lines) and after (dashed lines) the displacement. The 
imprint was generated by the pulse sequence depicted in Fig.~\ref{pulses}, 
the snapshots were taken at times $t/\tau_a=5$ for the initial imprint and 
$t/\tau_a=25$ for the displaced one.}
\end{figure}

For the second step we start with a $2\pi$-control pulse of duration $\tau_b$ 
decoupled from the medium. When this control pulse collides with the imprint, 
the signal pulse is retrieved which, upon interaction with the medium, stores 
its information in a displaced location. When the re-encoding has taken 
place, the control pulse recovers its original pulse area of $2\pi$ and 
propagates away at the speed of light. The end result is the displacement of 
the imprint farther into the medium with a $\pi$-phase shift 
for $\rho_{12}$ if $\tau_b<\tau_a$. The displacement is controlled by the 
phase-lag parameter defined as:
\begin{equation}
\delta^{ab}=\kappa_a x_2 -\kappa_a x_1=\ln \left|\frac{\tau_a+\tau_b}
{\tau_a-\tau_b}\right|,
\label{dis}
\end{equation}
where $x_2$ is the new location of the imprint. Note that the addition of 
Doppler broadening would affect the definition of the absorption coefficient 
and thus change the group velocity of the pulses in the medium as was shown 
in \cite{clader2007two}, but the storage procedure would carry through. The 
results of creation and displacement of the imprint are shown in 
Fig.~\ref{density}. Continuous lines show the first imprinted density matrix 
elements, and the dashed lines show their displacements.

Having reviewed the main results of the analytical solution we now address 
the question of how these ideas can be used for storage and retrieval of 
optical pulses. The first step is to consider media of finite length and use 
Eq.~\eqref{ximp} to determine the pulses' input areas in order to store the 
signal pulse at the desired location. This location must be chosen so that 
most of the imprint fits inside the medium for optimal information storage. 
Signal pulse storage takes place as described by the dynamics of the first 
step in the sequence of pulses, where the control pulse overtakes the signal 
pulse. Now that we have the information stored inside the medium we want to 
be able to retrieve it. To do this we inject a second control pulse of area 
$2\pi$ and duration such that according to Eq.~\eqref{dis} the imprint is 
pushed outside the medium. When the control pulse collides with the imprint 
it retrieves the signal pulse that was stored in the medium, as described by 
the second step in the sequence. According to the previous results, the 
signal pulse travels until it reaches the location where the imprint is 
supposed to be re-encoded. However, this never takes place because the 
location lies outside the medium. Thus by frustrating the signal pulse 
re-storage by means of the end face of the medium we are able to retrieve it. 
Of course this retrieval can be done in any number of steps, displacing the 
imprint closer to the end face before retrieving it. 

\section{Numerical Results}

\subsection{Initial considerations}

The analytical solution is fairly restrictive because it assumes an infinite 
medium and $2\pi$ pulses with asymptotic hyperbolic secant shape with 
infinitely long tails. Additionally, we have neglected the effects of 
homogeneous relaxation phenomena and the atom-field coupling parameters were 
kept equal. Since we are considering the propagation of pulses in an ultra-cold 
atomic system, we can safely omit the effects of collisional and Doppler 
broadening, but spontaneous emission $\Gamma_3$ is still present and could 
have a noticeable effect. We modify the von Neumann equation \eqref{neumann2} 
to account for this:
\begin{subequations}
\begin{align}
\pd{\rho_{11}}{T}=&\frac{i}{2}\Omega_{13}^*\rho_{31}-\frac{i}
{2}\Omega_{13}\rho_{13}+\frac{\Gamma_3}{2}\rho_{33}\\
\pd{\rho_{22}}{T}=&\frac{i}{2}\Omega_{23}^*\rho_{32}-\frac{i}
{2}\Omega_{23}\rho_{23}+\frac{\Gamma_3}{2}\rho_{33}\\
\pd{\rho_{33}}{T}=&-\frac{i}{2}\Omega_{13}^*\rho_{31}+\frac{i}
{2}\Omega_{13}\rho_{13}-\frac{i}{2}\Omega_{23}^*\rho_{32}+\frac{i}
{2}\Omega_{23}\rho_{23} \nonumber\\
&-\Gamma_3\rho_{33}\\
\pd{\rho_{12}}{T}=&\frac{i}{2}\Omega_{13}^*\rho_{32}-\frac{i}
{2}\Omega_{23}\rho_{13}\\
\pd{\rho_{13}}{T}=&i\Delta\rho_{13}-\frac{i}
{2}\Omega_{23}^*\rho_{12}+\frac{i}
{2}\Omega_{13}^*(\rho_{33}-\rho_{11})-\frac{\Gamma_3}{2}\rho_{13}\\
\pd{\rho_{23}}{T}=&i\Delta\rho_{23}-\frac{i}
{2}\Omega_{13}^*\rho_{21}+\frac{i}
{2}\Omega_{23}^*(\rho_{33}-\rho_{22})-\frac{\Gamma_3}{2}\rho_{23}
\end{align}
\end{subequations}
and for the field:
\begin{subequations}
\begin{align}
\pd{\Omega_{13}}{Z}&=i\mu_{13} \rho_{31}\\
\pd{\Omega_{23}}{Z}&=i\mu_{23} \rho_{32}.
\end{align}
\label{meqsn}
\end{subequations}
 
The time and length scales are respectively defined in terms of the duration 
and absorption coefficient associated with the first signal pulse. We abandon 
the idealized conditions of infinitely long media and pulses by considering a 
medium ten absorption lengths long (unless otherwise noted) and setting the 
Rabi frequencies to zero when $|\tau_a\Omega|<10^{-5}$. For each simulation 
we will consider three cases: hyperbolic-secant-shaped pulses with no decay 
channels, Gaussian-shaped pulses with no decay channels, and hyperbolic-secant 
shaped pulses with decay channels. This allows the effects of shape and 
homogeneous broadening to be studied separately. The shape of the pulses we 
will use throughout are:
\begin{subequations}
\begin{align}
\Omega=\frac{\theta}{\pi\tau}\text{sech}\left(\frac{T}{\tau}\right)\\
\shortintertext{and}
\Omega= \frac{\theta}{\sqrt{2\pi}\tau}e^{-\frac{T^2}{2\tau^2}}.
\end{align}
\end{subequations}

For the simulations, we consider a sample of $^{87}$Rb using the $D_2$ line 
and consider a pulse duration such that $\tau_a \Gamma_3=0.01$, so 
$\tau_a\approx 0.26\,\text{ns}$. For the atom-field coupling parameters we 
have that $\mu_{23}/\mu_{13}=0.99998$ \cite{steck2001rubidium}. From these 
estimates we clearly see that the approximations made in order to get the 
analytical solution were justified. Nonetheless we need to study their 
effects to determine whether these pulse dynamics are an experimentally 
realizable scenario. For simplicity the detuning is taken to be zero. 
Another thing worth noting is that we cannot choose an 
arbitrary pulse duration because we need to be able to resolve the hyperfine 
splitting of the ground state but not of the excited state in order to have a 
$\Lambda$-system. For the case of Rb we have the condition that $0.15\,
\text{ns}<\tau_a<2\,\text{ns}$. We could have considered Cs atoms to attain 
smaller values for $\tau_a \Gamma_3$ by using shorter pulses, down to $0.1\,
\text{ns}$ \cite{steck2003cesium}. Or if we want to use longer pulses, we 
could use K but with the compromise of a larger value for $\tau_a 
\Gamma_3$($>0.03$) \cite{tiecke2010properties}. Note that the important 
quantity is $\tau_a \Gamma_3$ and not just $\Gamma_3$, because
 what matters here is 
how long the spontaneous decay time is with respect to the interaction time 
between the pulses and the atomic system.

\subsection{Location of initial imprint}

\begin{figure*}
\includegraphics[width=.8\linewidth]{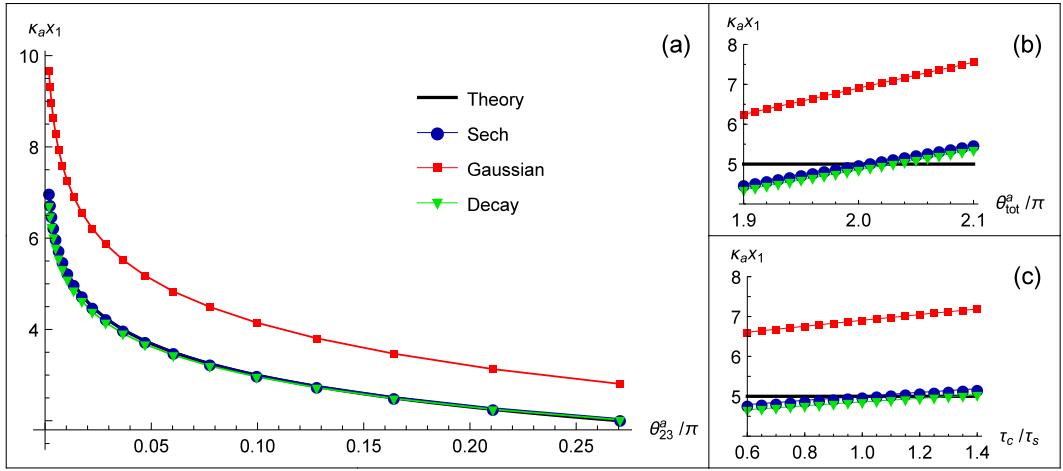}
\caption{\label{figx1} (Color online) Location of the initial imprint: 
(a) as a function of 
the control pulse area with a $2\pi$ signal pulse, (b) as a function of the 
total pulse area keeping the ratio of the pulses' areas constant so that the 
predicted location is $\kappa_a x_1=5$, and (c) as a function of the duration 
of the control pulse with a $2\pi$ signal pulse and a predicted location of 
$\kappa_a x_1=5$. These plots compare the location of the initial imprint for
four different cases. The black solid line is the formula given by the 
analytical solution, Eq.~\eqref{ximp}. The three plots with the markers 
represent the results of the numerical simulations for which we considered a 
medium ten absorption lengths long, finite pulses taking $\Omega=0$ when 
$|\tau_a\Omega|<10^{-5}$, and an initial $2\pi$ signal pulse. The plot with 
blue circle markers represents the case of a hyperbolic-secant pulse shape 
with $\tau_a \Gamma_3=0$, the one with green upside down 
triangle markers adds the effects 
of spontaneous emission to the previous one with $\tau_a \Gamma_3=0.01$, and 
finally the plot with red square markers considers a Gaussian-shaped pulse 
with $\tau_a \Gamma_3=0$.}
\end{figure*}

First, we consider how the shape and finiteness of the pulses, as well as the 
effects of homogeneous relaxation, affect the location of the first imprint. 
The results of the numerical simulations are summarized in Fig.~\ref{figx1}. 
For the dependence on the control pulse area, Fig.~\ref{figx1}(a), we notice 
that the hyperbolic secant-shaped pulse completely overlaps the curve given 
by the theory. We see that the effect of spontaneous emission is to lower the 
curve very slightly while maintaining the same shape. The Gaussian-shaped 
pulse clearly deviates from the expected behavior, but keeps a predictable 
trend and in the grand scheme has the same dependence, i.e., the location of 
the imprint increases when the control pulse area decreases.
The effects of spontaneous emission on the Gaussian-shaped pulse curve are 
analogous to  those of the hyperbolic secant-shaped pulse, namely, to lower 
the curve while keeping the same type of dependence (the data for this case 
are not presented for the sake of clarity in the figures).

The analytical solution sets the total pulse area of each step of the process 
equal to $2\pi$, but we are free to change that in the simulations. From an 
experimental point of view, it is important to know this dependence, since it 
might be difficult to control the area of the pulses with much precision. 
Setting the ratio of the signal to control pulse area such that 
Eq.~\eqref{ximp} predicts an imprint location of $\kappa_a x_1=5$, we vary 
the total pulse area [see Fig.~\ref{figx1}(b)]. We find that for all three 
cases there appears to be a linear dependence with similar positive slope, 
$\Delta (\kappa_a x_1)/\Delta \theta_\text{tot}^a\approx 5/\pi$ for the 
hyperbolic secant shaped- and $\Delta (\kappa_a 
x_1)/\Delta \theta_\text{tot}^a \approx 6.5/\pi$ for the Gaussian-shaped 
pulses. Here we only consider small variations inherent to any realistic 
experimental scenario and so this behavior cannot be extrapolated to 
arbitrarily large and small pulse areas. In particular, if we consider areas 
smaller than $\pi$ then the pulses are not strong enough to promote the 
necessary population transfer for the initial SIT propagation and then the 
transfer from signal to control pulses. For pulse areas larger than $3\pi$ we 
would get pulse breakup and thus a different kind of interaction.

Another restriction 
from the solution is that signal and control pulses are time matched (this 
might not be ideal for an experiment because one would have to change the duration 
of the control pulses between consecutive steps). Relaxing this condition we 
find a behavior similar to the previous case [see Fig.~\ref{figx1}(c)]: the 
imprint location increases with the duration of the control pulse. Once again 
all three cases seem to have similar positive slopes, $\Delta 
(\kappa_a x_1)/\Delta (\tau_{c}/\tau_s)\approx 0.5$ for the hyperbolic 
secants and $\Delta (\kappa_a x_1)/\Delta 
(\tau_{c}/\tau_s) \approx 0.73$ for 
the Gaussian. It is also worth mentioning that the shape of the imprint is 
the same as the one predicted theoretically (Fig.~\ref{density}) for the two 
cases where the pulses are hyperbolic secants, but for the Gaussian-shaped 
pulse it is approximately $1.4$ times wider. This is a consequence of the 
reshaping of the Gaussian pulse to a corresponding hyperbolic-secant-shaped 
pulse with slightly different time duration.

\begin{figure*}
\includegraphics[width=.8\linewidth]{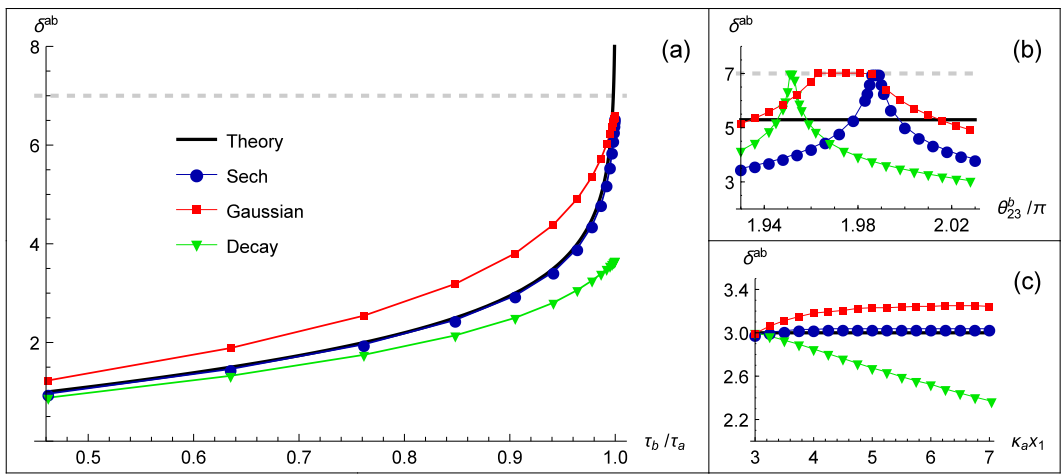}
\caption{\label{figdel} (Color online) Displacement of the 
imprint as a function of: (a) the 
duration of the second control pulse, (b) the control pulse area, and (c) the 
location of the initial imprint. We consider the same three cases as those 
described in Fig.~\ref{figx1}, represented by the same markers. For (a) and 
(b) we considered an initial $2\pi$ signal pulse with the necessary pulse 
area of the corresponding control pulse so that the location of the first 
imprint is at $\kappa_a x_1=3$. Then for (a) we kept the area of the second 
control pulse equal to $2\pi$. The dashed light gray line 
marks the end of the medium 
and the black solid line is the result predicted by Eq.~\eqref{dis}. For (c) 
we considered a longer medium, 15 absorption lengths long, and we chose the 
control pulse duration so that we had a displacement of $\delta^{ab}=3$ when 
the initial imprint was at $\kappa_a x_1=3$. }
\end{figure*}

\subsection{\label{secdis} Displacement of the imprint}

Now that we know we can imprint the information of the signal pulse into the 
atomic system given non-idealized conditions, we have to study the next step 
of the pulse sequence: the displacement of the imprint. To do this, 
we select the initial pulse area of the control pulse for each case using the 
results from the previous section, so that the initial imprint is made at 
$\kappa_a x_1=3$ combined with a $2\pi$ signal pulse. Then we vary the 
control pulse duration and find the new position of the imprint. The results 
are summarized in Fig.~\ref{figdel}(a). 

As predicted by the theory, the closer $\tau_b$ is to $\tau_a$, the more the 
imprint is displaced. Additionally, the new imprint is identical to the 
initial one except for a $\pi$ phase shift when the control pulse duration is 
smaller than the signal pulse. We also notice that the different parameters 
affect the displacement in a similar way regarding the location of the 
initial imprint. 
The hyperbolic-secant-shaped pulse closely follows the behavior dictated by 
Eq.~\eqref{dis}, while the addition of decay causes a decrease in the 
displacement. However, the Gaussian-shaped pulse causes a displacement that 
is typically larger than that given by the analytical solution (here again 
the effects of spontaneous emission for the Gaussian-shaped pulse are 
analogous to those of the hyperbolic-secant-shaped pulse). The biggest 
difference from the theoretical prediction is that there is an upper limit in 
how much the imprint can be moved (this is true for all three cases). This 
will define how close the imprint must be to the end face so that the signal 
pulse can be retrieved. Another 
feature is that if we consider the case $\tau_b>\tau_a$ we get similar 
results but the imprint does not present the $\pi$ phase shift. This might be 
desirable if we want the retrieved signal pulse to have the same phase as the 
original (as we will see in the next section), but one must be careful 
because homogeneous relaxation will affect the displacement even more if the 
pulses interact with the medium for a longer time.  

We also explore the dependence of the displacement with respect to the 
control pulse area $\theta_{23}^b$ as noted before, a parameter not 
accessible from the analytical solution. As shown in Fig.~\ref{figdel}(b), the 
results are quite surprising. We find that the displacement increases as we 
decrease the pulse area, until it reaches a maximum (these are hidden by the 
plateaus which represent that the imprint has been pushed outside the medium) 
and then starts decreasing again. For all three cases, the ``optimal'' pulse 
area is actually less than the $2\pi$ predicted by the theory, and not only 
do we have maxima for these smaller areas but also the 
displacement can be much 
greater than the one predicted in Eq.~\eqref{dis}.

There is also some dependence on the first imprint location, and there are two 
possible tendencies as shown in Fig.~\ref{figdel}(c). The first is an increase 
in the displacement as we increase $x_1$ until it reaches a steady value. 
This behavior is obtained when there are no decay channels and is due to the 
finiteness of the medium which cuts off part of the information deposited by 
the signal pulse, so the closer we get to the center of the medium the more 
room there is for the imprint to be made. As for the Gaussian-shaped pulse, 
it is affected the most because the imprint is wider. When we consider 
spontaneous emission, another process takes over: The further into the medium 
the first imprint is made, the less it will be displaced by the second 
control pulse. This can be understood by the fact that the longer it takes to 
deposit the information of the signal pulse, the longer the decay is 
effective, provoking some information loss and this ultimately leads to less 
displacement.

\subsection{Retrieval of the signal pulse}

One of the most important effects of the boundaries in the medium is to cut 
off the pulse transfer process of the analytical solution, thus providing a 
way to retrieve the initial signal pulse. To quantify the accuracy 
of the retrieval process we define the retrieval efficiency as
\begin{equation}
\eta=\frac{I_{13}^{\text{out}}}{I_{13}^{\text{in}}}=\frac{\int|
\Omega_{13}^{\text{out}}|^2 dt}{\int|\Omega_{13}^{\text{in}}|^2dt}.
\label{fidel}
\end{equation}
Clearly this quantity only gives information about the output signal pulse 
intensity but does not take into account any possible reshaping of the pulse. 
To account for this we calculate the correlation coefficient $r$ between the 
input and output signal pulses.
This is particularly important for the Gaussian-shaped pulses which are 
reshaped into hyperbolic secants as they propagate through the medium. This 
has already been noted in the case of atomic vapors at room temperature 
\cite{clader2007two,clader2008two}. From the previous sections it should be 
clear that the retrieval process can take place in any number of steps, but 
for the sake of clarity we will treat only two cases: two- and three-step 
retrieval. Additionally, we will only consider control pulses of areas equal 
to $2\pi$ and of duration times  equal to or smaller than that of the initial 
signal pulse for the displacement and retrieval steps.

{
\renewcommand{\arraystretch}{1.3}
\setlength{\tabcolsep}{5pt}
\begin{table}
\caption{\label{tab} Retrieval efficiency and correlation coefficient for one- 
and two-step processes.}
\begin{tabular}{lcccc}
\hline
\hline 
Case & \multicolumn{2}{c}{1 step} & \multicolumn{2}{c}{2 steps}\\
& $\eta$ & $r$ & $\eta$ & $r$\\
 \hline
 Sech& 98\%& 1.0000& 98\%& 1.0000\\ 
 Decay& 65\%& 0.9998& 66\%& 0.9998\\ 
 Gaussian& 94\%&0.9940 & 94\%& 0.9941 \\
 \hline
 \hline
\end{tabular}
\end{table}
}
  
Let us first consider the one-step process. Here we want to make the imprint 
somewhere in the medium such that it can be retrieved without having to move 
it closer to the boundary (this limit is set by the case where we consider 
homogeneous relaxation). For this we consider the necessary pulse areas for 
each case so that the initial imprint is made at $\kappa_a x_1=8$, and for 
the second step we consider a control pulse with the same duration as the 
signal pulse, giving us the maximum displacement. For the two-step process we 
will make the initial imprint at $\kappa_a x_1=5$, then tailor the time 
duration of the following control pulse so that we displace the imprint by 
$\delta^{ab}=3$ and finally push the imprint outside the medium by means of 
another control pulse of duration equal to the original signal pulse. The 
results are summarized in Table \ref{tab}. We notice that, when no decay is 
present, the efficiency is high (larger than 90\%) but as soon as homogeneous 
relaxation is added it drops to 65\% for one-step and 66\% 
for two-step processes. As 
for the correlation coefficient, it is fairly close to unity in all cases, 
indicating that the shape of the pulse is mostly preserved throughout the 
storage and retrieval procedure. The Gaussian-shaped pulses are the ones that 
have the lowest $r$ values as could have been expected because of the 
reshaping during propagation. In 
any case, we are able to retrieve a good portion of the initial signal pulse 
and see that this does not depend on the number of steps involved. In the 
one-step case, the retrieved signal pulse has a $\pi$ phase shift with 
respect to the original. The correct phase can be obtained by inverting the 
initial storing control pulse or by increasing the duration  of the second 
control pulse as discussed in Sec.~\ref{secdis}. In the two-step process, the 
signal pulse comes out with the same phase due to the $\pi$-phase shift in 
the displacement step.

\section{Conclusions}

In this report we have shown that previously predicted memory manipulation by 
means of idealized atomic pulse dynamics is plausible even in non-idealized 
conditions. The shape of the pulse and the effects of spontaneous emission 
have an impact on the quantitative results but the storage and retrieval of 
the signal pulse are achieved in all cases. We have quantified the deviations 
in each case and even shown some features that could add more control to 
the process and lift some restrictions. The length of the medium has no 
effect on the memory manipulation, but should be chosen so that most of the 
information can be deposited into the medium. This may no longer be true if 
we consider the reflection of the pulses at the end face.

Another aspect that we noticed throughout this work, and that has been noted 
elsewhere, is the stability against the total pulse area. When considering 
pulses of different area than $2\pi$, be it bigger or smaller, the pulses are 
reshaped as they propagate in an effort to obtain a $2\pi$ total area. This 
is completely analogous to the predictions of the area theorem 
\cite{mccall1967self,mccall1969self} for a homogeneously broadened two-level 
atom. Additionally, we can ask which is the appropriate extension of the 
total pulse area for higher order solutions, particularly when the two first-
order solutions overlap.

We can also think of extending this analysis to more complicated pulse 
sequences. This would be the same as simulating higher-order solutions of the 
Maxwell-Bloch equations obtained by further applications of the nonlinear 
superposition rule. This could possibly pave the way for making multiple 
imprints belonging to different signal pulses and then manipulating them by 
means of control pulses.

\begin{acknowledgments}
This work was supported by  NSF grant PHY-1203931, and and R.G.-C. acknowledges support as a 
CONACYT Fellow.  
\end{acknowledgments}


\begin{thebibliography}{25}%
\makeatletter
\providecommand \@ifxundefined [1]{%
 \@ifx{#1\undefined}
}%
\providecommand \@ifnum [1]{%
 \ifnum #1\expandafter \@firstoftwo
 \else \expandafter \@secondoftwo
 \fi
}%
\providecommand \@ifx [1]{%
 \ifx #1\expandafter \@firstoftwo
 \else \expandafter \@secondoftwo
 \fi
}%
\providecommand \natexlab [1]{#1}%
\providecommand \enquote  [1]{``#1''}%
\providecommand \bibnamefont  [1]{#1}%
\providecommand \bibfnamefont [1]{#1}%
\providecommand \citenamefont [1]{#1}%
\providecommand \href@noop [0]{\@secondoftwo}%
\providecommand \href [0]{\begingroup \@sanitize@url \@href}%
\providecommand \@href[1]{\@@startlink{#1}\@@href}%
\providecommand \@@href[1]{\endgroup#1\@@endlink}%
\providecommand \@sanitize@url [0]{\catcode `\\12\catcode `\$12\catcode
  `\&12\catcode `\#12\catcode `\^12\catcode `\_12\catcode `\%12\relax}%
\providecommand \@@startlink[1]{}%
\providecommand \@@endlink[0]{}%
\providecommand \url  [0]{\begingroup\@sanitize@url \@url }%
\providecommand \@url [1]{\endgroup\@href {#1}{\urlprefix }}%
\providecommand \urlprefix  [0]{URL }%
\providecommand \Eprint [0]{\href }%
\providecommand \doibase [0]{http://dx.doi.org/}%
\providecommand \selectlanguage [0]{\@gobble}%
\providecommand \bibinfo  [0]{\@secondoftwo}%
\providecommand \bibfield  [0]{\@secondoftwo}%
\providecommand \translation [1]{[#1]}%
\providecommand \BibitemOpen [0]{}%
\providecommand \bibitemStop [0]{}%
\providecommand \bibitemNoStop [0]{.\EOS\space}%
\providecommand \EOS [0]{\spacefactor3000\relax}%
\providecommand \BibitemShut  [1]{\csname bibitem#1\endcsname}%
\let\auto@bib@innerbib\@empty
\bibitem [{\citenamefont {McCall}\ and\ \citenamefont
  {Hahn}(1967)}]{mccall1967self}%
  \BibitemOpen
  \bibfield  {author} {\bibinfo {author} {\bibfnamefont {S.~L.}\ \bibnamefont
  {McCall}}\ and\ \bibinfo {author} {\bibfnamefont {E.~L.}\ \bibnamefont
  {Hahn}},\ }\href {\doibase 10.1103/PhysRevLett.18.908} {\bibfield  {journal}
  {\bibinfo  {journal} {Phys. Rev. Lett.}\ }\textbf {\bibinfo {volume} {18}},\
  \bibinfo {pages} {908} (\bibinfo {year} {1967})}\BibitemShut {NoStop}%
\bibitem [{\citenamefont {McCall}\ and\ \citenamefont
  {Hahn}(1969)}]{mccall1969self}%
  \BibitemOpen
  \bibfield  {author} {\bibinfo {author} {\bibfnamefont {S.~L.}\ \bibnamefont
  {McCall}}\ and\ \bibinfo {author} {\bibfnamefont {E.~L.}\ \bibnamefont
  {Hahn}},\ }\href {\doibase 10.1103/PhysRev.183.457} {\bibfield  {journal}
  {\bibinfo  {journal} {Phys. Rev.}\ }\textbf {\bibinfo {volume} {183}},\
  \bibinfo {pages} {457} (\bibinfo {year} {1969})}\BibitemShut {NoStop}%
\bibitem [{\citenamefont {Gray}\ \emph {et~al.}(1978)\citenamefont {Gray},
  \citenamefont {Whitley},\ and\ \citenamefont {Stroud}}]{gray1978coherent}%
  \BibitemOpen
  \bibfield  {author} {\bibinfo {author} {\bibfnamefont {H.~R.}\ \bibnamefont
  {Gray}}, \bibinfo {author} {\bibfnamefont {R.~M.}\ \bibnamefont {Whitley}}, \
  and\ \bibinfo {author} {\bibfnamefont {C.~R.}\ \bibnamefont {Stroud}},\
  }\href {\doibase 10.1364/OL.3.000218} {\bibfield  {journal} {\bibinfo
  {journal} {Opt. Lett.}\ }\textbf {\bibinfo {volume} {3}},\ \bibinfo {pages}
  {218} (\bibinfo {year} {1978})}\BibitemShut {NoStop}%
\bibitem [{\citenamefont {Kasapi}\ \emph {et~al.}(1995)\citenamefont {Kasapi},
  \citenamefont {Jain}, \citenamefont {Yin},\ and\ \citenamefont
  {Harris}}]{kasapi1995electromagnetically}%
  \BibitemOpen
  \bibfield  {author} {\bibinfo {author} {\bibfnamefont {A.}~\bibnamefont
  {Kasapi}}, \bibinfo {author} {\bibfnamefont {M.}~\bibnamefont {Jain}},
  \bibinfo {author} {\bibfnamefont {G.~Y.}\ \bibnamefont {Yin}}, \ and\
  \bibinfo {author} {\bibfnamefont {S.~E.}\ \bibnamefont {Harris}},\ }\href
  {\doibase 10.1103/PhysRevLett.74.2447} {\bibfield  {journal} {\bibinfo
  {journal} {Phys. Rev. Lett.}\ }\textbf {\bibinfo {volume} {74}},\ \bibinfo
  {pages} {2447} (\bibinfo {year} {1995})}\BibitemShut {NoStop}%
\bibitem [{\citenamefont {Harris}(1997)}]{harris2008electromagnetically}%
  \BibitemOpen
  \bibfield  {author} {\bibinfo {author} {\bibfnamefont {S.~E.}\ \bibnamefont
  {Harris}},\ }\href@noop {} {\bibfield  {journal} {\bibinfo  {journal} {Phys.
  Today}\ }\textbf {\bibinfo {volume} {50}},\ \bibinfo {pages} {36} (\bibinfo
  {year} {1997})}\BibitemShut {NoStop}%
\bibitem [{\citenamefont {Garrett}\ and\ \citenamefont
  {McCumber}(1970)}]{garrett1970propagation}%
  \BibitemOpen
  \bibfield  {author} {\bibinfo {author} {\bibfnamefont {C.~G.~B.}\
  \bibnamefont {Garrett}}\ and\ \bibinfo {author} {\bibfnamefont {D.~E.}\
  \bibnamefont {McCumber}},\ }\href {\doibase 10.1103/PhysRevA.1.305}
  {\bibfield  {journal} {\bibinfo  {journal} {Phys. Rev. A}\ }\textbf {\bibinfo
  {volume} {1}},\ \bibinfo {pages} {305} (\bibinfo {year} {1970})}\BibitemShut
  {NoStop}%
\bibitem [{\citenamefont {Boyd}\ and\ \citenamefont
  {Gauthier}(2002)}]{boyd2002slow}%
  \BibitemOpen
  \bibfield  {author} {\bibinfo {author} {\bibfnamefont {R.~W.}\ \bibnamefont
  {Boyd}}\ and\ \bibinfo {author} {\bibfnamefont {D.~J.}\ \bibnamefont
  {Gauthier}},\ }\href {\doibase
  http://dx.doi.org/10.1016/S0079-6638(02)80030-0} {\bibfield  {journal}
  {\bibinfo  {journal} {Prog. Opt.}\ }\textbf {\bibinfo {volume} {43}},\
  \bibinfo {pages} {497} (\bibinfo {year} {2002})}\BibitemShut {NoStop}%
\bibitem [{\citenamefont {Milonni}(2005)}]{milonni2004fast}%
  \BibitemOpen
  \bibfield  {author} {\bibinfo {author} {\bibfnamefont {P.~W.}\ \bibnamefont
  {Milonni}},\ }\href@noop {} {\emph {\bibinfo {title} {Fast Light, Slow Light
  and Left-Handed Light}}}\ (\bibinfo  {publisher} {Institute of Physics},\
  \bibinfo {address} {Bristol},\ \bibinfo {year} {2005})\BibitemShut {NoStop}%
\bibitem [{\citenamefont {Milonni}(1976)}]{milonni1976semiclassical}%
  \BibitemOpen
  \bibfield  {author} {\bibinfo {author} {\bibfnamefont {P.~W.}\ \bibnamefont
  {Milonni}},\ }\href {\doibase http://dx.doi.org/10.1016/0370-1573(76)90037-5}
  {\bibfield  {journal} {\bibinfo  {journal} {Phys. Rep.}\ }\textbf {\bibinfo
  {volume} {25}},\ \bibinfo {pages} {1} (\bibinfo {year} {1976})}\BibitemShut
  {NoStop}%
\bibitem [{\citenamefont {Grobe}\ \emph {et~al.}(1994)\citenamefont {Grobe},
  \citenamefont {Hioe},\ and\ \citenamefont {Eberly}}]{grobe1994formation}%
  \BibitemOpen
  \bibfield  {author} {\bibinfo {author} {\bibfnamefont {R.}~\bibnamefont
  {Grobe}}, \bibinfo {author} {\bibfnamefont {F.~T.}\ \bibnamefont {Hioe}}, \
  and\ \bibinfo {author} {\bibfnamefont {J.~H.}\ \bibnamefont {Eberly}},\
  }\href {\doibase 10.1103/PhysRevLett.73.3183} {\bibfield  {journal} {\bibinfo
   {journal} {Phys. Rev. Lett.}\ }\textbf {\bibinfo {volume} {73}},\ \bibinfo
  {pages} {3183} (\bibinfo {year} {1994})}\BibitemShut {NoStop}%
\bibitem [{\citenamefont {Fleischhauer}\ and\ \citenamefont
  {Lukin}(2000)}]{fleischhauer2000dark}%
  \BibitemOpen
  \bibfield  {author} {\bibinfo {author} {\bibfnamefont {M.}~\bibnamefont
  {Fleischhauer}}\ and\ \bibinfo {author} {\bibfnamefont {M.~D.}\ \bibnamefont
  {Lukin}},\ }\href {\doibase 10.1103/PhysRevLett.84.5094} {\bibfield
  {journal} {\bibinfo  {journal} {Phys. Rev. Lett.}\ }\textbf {\bibinfo
  {volume} {84}},\ \bibinfo {pages} {5094} (\bibinfo {year}
  {2000})}\BibitemShut {NoStop}%
\bibitem [{\citenamefont {Liu}\ \emph {et~al.}(2001)\citenamefont {Liu},
  \citenamefont {Dutton}, \citenamefont {Behroozi},\ and\ \citenamefont
  {Hau}}]{liu2001observation}%
  \BibitemOpen
  \bibfield  {author} {\bibinfo {author} {\bibfnamefont {C.}~\bibnamefont
  {Liu}}, \bibinfo {author} {\bibfnamefont {Z.}~\bibnamefont {Dutton}},
  \bibinfo {author} {\bibfnamefont {C.~H.}\ \bibnamefont {Behroozi}}, \ and\
  \bibinfo {author} {\bibfnamefont {L.~V.}\ \bibnamefont {Hau}},\ }\href@noop
  {} {\bibfield  {journal} {\bibinfo  {journal} {Nature (London)}\ }\textbf
  {\bibinfo {volume} {409}},\ \bibinfo {pages} {490} (\bibinfo {year}
  {2001})}\BibitemShut {NoStop}%
\bibitem [{\citenamefont {Camacho}\ \emph {et~al.}(2009)\citenamefont
  {Camacho}, \citenamefont {Vudyasetu},\ and\ \citenamefont
  {Howell}}]{camacho2009four}%
  \BibitemOpen
  \bibfield  {author} {\bibinfo {author} {\bibfnamefont {R.~M.}\ \bibnamefont
  {Camacho}}, \bibinfo {author} {\bibfnamefont {P.~K.}\ \bibnamefont
  {Vudyasetu}}, \ and\ \bibinfo {author} {\bibfnamefont {J.~C.}\ \bibnamefont
  {Howell}},\ }\href@noop {} {\bibfield  {journal} {\bibinfo  {journal} {Nature
  Photonics}\ }\textbf {\bibinfo {volume} {3}},\ \bibinfo {pages} {103}
  (\bibinfo {year} {2009})}\BibitemShut {NoStop}%
\bibitem [{\citenamefont {Groves}\ \emph {et~al.}(2013)\citenamefont {Groves},
  \citenamefont {Clader},\ and\ \citenamefont {Eberly}}]{groves2013jaynes}%
  \BibitemOpen
  \bibfield  {author} {\bibinfo {author} {\bibfnamefont {E.}~\bibnamefont
  {Groves}}, \bibinfo {author} {\bibfnamefont {B.~D.}\ \bibnamefont {Clader}},
  \ and\ \bibinfo {author} {\bibfnamefont {J.~H.}\ \bibnamefont {Eberly}},\
  }\href@noop {} {\bibfield  {journal} {\bibinfo  {journal} {J. Phys. B: At.
  Mol. Opt. Phys.}\ }\textbf {\bibinfo {volume} {46}},\ \bibinfo {pages}
  {224005} (\bibinfo {year} {2013})}\BibitemShut {NoStop}%
\bibitem [{\citenamefont {Allen}\ and\ \citenamefont
  {Eberly}(1987)}]{allen2012optical}%
  \BibitemOpen
  \bibfield  {author} {\bibinfo {author} {\bibfnamefont {L.}~\bibnamefont
  {Allen}}\ and\ \bibinfo {author} {\bibfnamefont {J.~H.}\ \bibnamefont
  {Eberly}},\ }\href@noop {} {\emph {\bibinfo {title} {Optical Resonance and
  Two-Level Atoms}}}\ (\bibinfo  {publisher} {Dover},\ \bibinfo {address} {New
  York},\ \bibinfo {year} {1987})\BibitemShut {NoStop}%
\bibitem [{\citenamefont {Park}\ and\ \citenamefont
  {Shin}(1998)}]{park1998matched}%
  \BibitemOpen
  \bibfield  {author} {\bibinfo {author} {\bibfnamefont {{Q.~Han}}\ \bibnamefont
  {Park}}\ and\ \bibinfo {author} {\bibfnamefont {H.~J.}\ \bibnamefont
  {Shin}},\ }\href {\doibase 10.1103/PhysRevA.57.4643} {\bibfield  {journal}
  {\bibinfo  {journal} {Phys. Rev. A}\ }\textbf {\bibinfo {volume} {57}},\
  \bibinfo {pages} {4643} (\bibinfo {year} {1998})}\BibitemShut {NoStop}%
\bibitem [{\citenamefont {Clader}\ and\ \citenamefont
  {Eberly}(2007)}]{clader2007two}%
  \BibitemOpen
  \bibfield  {author} {\bibinfo {author} {\bibfnamefont {B.~D.}\ \bibnamefont
  {Clader}}\ and\ \bibinfo {author} {\bibfnamefont {J.~H.}\ \bibnamefont
  {Eberly}},\ }\href {\doibase 10.1103/PhysRevA.76.053812} {\bibfield
  {journal} {\bibinfo  {journal} {Phys. Rev. A}\ }\textbf {\bibinfo {volume}
  {76}},\ \bibinfo {pages} {053812} (\bibinfo {year} {2007})}\BibitemShut
  {NoStop}%
\bibitem [{\citenamefont {Lamb}(1980)}]{lamb1980elements}%
  \BibitemOpen
  \bibfield  {author} {\bibinfo {author} {\bibfnamefont {G.~L.}\ \bibnamefont
  {Lamb}},\ }\href@noop {} {\emph {\bibinfo {title} {Elements of Soliton
  Theory}}}\ (\bibinfo  {publisher} {Wiley},\ \bibinfo {address} {New York},\
  \bibinfo {year} {1980})\BibitemShut {NoStop}%
\bibitem [{\citenamefont {Miura}(1976)}]{miura1976backlund}%
  \BibitemOpen
  \bibfield  {author} {\bibinfo {author} {\bibfnamefont {R.~M.}\ \bibnamefont
  {Miura}},\ }\href@noop {} {\emph {\bibinfo {title} {Backlund
  Transformations}}}\ (\bibinfo  {publisher} {Springer-Verlag},\ \bibinfo
  {address} {Berlin},\ \bibinfo {year} {1976})\BibitemShut {NoStop}%
\bibitem [{\citenamefont {Gu}\ \emph {et~al.}(2005)\citenamefont {Gu},
  \citenamefont {Hu},\ and\ \citenamefont {Zhou}}]{gu2006darboux}%
  \BibitemOpen
  \bibfield  {author} {\bibinfo {author} {\bibfnamefont {C.}~\bibnamefont
  {Gu}}, \bibinfo {author} {\bibfnamefont {H.}~\bibnamefont {Hu}}, \ and\
  \bibinfo {author} {\bibfnamefont {Z.}~\bibnamefont {Zhou}},\ }\href@noop {}
  {\emph {\bibinfo {title} {Darboux Transformations in Integrable Systems}}}\
  (\bibinfo  {publisher} {Springer},\ \bibinfo {address} {Dordrecht},\ \bibinfo
  {year} {2005})\BibitemShut {NoStop}%
\bibitem [{\citenamefont {Cie{\'s}li{\'n}ski}(2009)}]{cieslinski2009algebraic}%
  \BibitemOpen
  \bibfield  {author} {\bibinfo {author} {\bibfnamefont {J.~L.}\ \bibnamefont
  {Cie{\'s}li{\'n}ski}},\ }\href@noop {} {\bibfield  {journal} {\bibinfo
  {journal} {J. Phys. A: Math. Theor.}\ }\textbf {\bibinfo {volume} {42}},\
  \bibinfo {pages} {404003} (\bibinfo {year} {2009})}\BibitemShut {NoStop}%
\bibitem [{\citenamefont {Steck}(2010{\natexlab{a}})}]{steck2001rubidium}%
  \BibitemOpen
  \bibfield  {author} {\bibinfo {author} {\bibfnamefont {D.~A.}\ \bibnamefont
  {Steck}},\ }\href@noop {} {\enquote {\bibinfo {title} {Rubidium 87 {D} {L}ine
  {D}ata},}\ }\bibinfo {howpublished} {available online at
  \url{http://steck.us/alkalidata}} (\bibinfo {year} {revision 2.1.4, 23
  December 2010}{\natexlab{a}})\BibitemShut {NoStop}%
\bibitem [{\citenamefont {Steck}(2010{\natexlab{b}})}]{steck2003cesium}%
  \BibitemOpen
  \bibfield  {author} {\bibinfo {author} {\bibfnamefont {D.~A.}\ \bibnamefont
  {Steck}},\ }\href@noop {} {\enquote {\bibinfo {title} {Cesium {D} {L}ine
  {D}ata},}\ }\bibinfo {howpublished} {available online at
  \url{http://steck.us/alkalidata}} (\bibinfo {year} {revision 2.1.4, 23
  December 2010}{\natexlab{b}})\BibitemShut {NoStop}%
\bibitem [{\citenamefont {Tiecke}(2011)}]{tiecke2010properties}%
  \BibitemOpen
  \bibfield  {author} {\bibinfo {author} {\bibfnamefont {T.~G.}\ \bibnamefont
  {Tiecke}},\ }\href
  {http://www.tobiastiecke.nl/archive/PotassiumProperties.pdf} {\enquote
  {\bibinfo {title} {Properties of {P}otassium},}\ }\bibinfo {howpublished}
  {available online at
  \url{http://www.tobiastiecke.nl/archive/PotassiumProperties.pdf}} (\bibinfo
  {year} {v1.02, May 2011})\BibitemShut {NoStop}%
\bibitem [{\citenamefont {Clader}\ and\ \citenamefont
  {Eberly}(2008)}]{clader2008two}%
  \BibitemOpen
  \bibfield  {author} {\bibinfo {author} {\bibfnamefont {B.~D.}\ \bibnamefont
  {Clader}}\ and\ \bibinfo {author} {\bibfnamefont {J.~H.}\ \bibnamefont
  {Eberly}},\ }\href {\doibase 10.1103/PhysRevA.78.033803} {\bibfield
  {journal} {\bibinfo  {journal} {Phys. Rev. A}\ }\textbf {\bibinfo {volume}
  {78}},\ \bibinfo {pages} {033803} (\bibinfo {year} {2008})}\BibitemShut
  {NoStop}%
\end{thebibliography}
\end{document}